\documentclass[twocolumn,showpacs,showkeys,amsmath,amssymb,floatfix]{revtex4}

\usepackage{graphicx}
\usepackage{bm} 
\usepackage[T1]{fontenc}

\def\vec#1{\mathchoice{\mbox{\boldmath$\displaystyle#1$}}
{\mbox{\boldmath$\textstyle#1$}}
{\mbox{\boldmath$\scriptstyle#1$}}
{\mbox{\boldmath$\scriptscriptstyle#1$}}}
\makeatletter
\newcommand\erfc{\mathop{\operator@font erfc}\nolimits}
\def\slashchar#1{\setbox0=\hbox{$#1$}
   \dimen0=\wd0 \setbox1=\hbox{/} \dimen1=\wd1
   \ifdim\dimen0>\dimen1 \rlap{\hbox to \dimen0{\hfil/\hfil}} #1
   \else  \rlap{\hbox to \dimen1{\hfil$#1$\hfil}} / \fi}

\makeatother

\begin{document}
 
\title{Femtoscopy in hydro-inspired models with resonances
\footnote{Research supported by the Polish Ministry of Education and
Science (former Polish State Committee for  
Scientific Research), grants 2P03B~05925, 2P03B 02828 and 1P03B~10929. 
Research carried out within the scope of the ERG (GDRE): {\it Heavy ions at
ultrarelativistic energies}  -- a European Research Group comprising
IN2P3/CNRS, Ecole des Mines de Nantes, Universite de Nantes, Warsaw
University of Technology, JINR Dubna, ITEP Moscow and Bogolyubov
Institute for Theoretical Physics NAS of Ukraine.
}}

\author{Adam Kisiel} 
\email{kisiel@if.pw.edu.pl}
\affiliation{Faculty of Physics, Warsaw University of Technology, PL-00661 Warsaw, Poland}
\affiliation{SUBATECH, Laboratoire de Physique Subatomique et des Technologies Associees, EMN-IN2P3/CNRS-Universite, Nantes, F-44307, France}

\author{Wojciech Florkowski} 
\email{Wojciech.Florkowski@ifj.edu.pl}
\affiliation{Institute of Physics, \'Swi\c{e}tokrzyska Academy,
ul.~\'Swi\c{e}tokrzyska 15, PL-25406~Kielce, Poland} 
\affiliation{The H. Niewodnicza\'nski Institute of Nuclear Physics, 
Polish Academy of Sciences, PL-31342 Krak\'ow, Poland}

\author{Wojciech Broniowski} 
\email{Wojciech.Broniowski@ifj.edu.pl} 
\affiliation{Institute of Physics, \'Swi\c{e}tokrzyska Academy,
ul.~\'Swi\c{e}tokrzyska 15, PL-25406~Kielce, Poland} 
\affiliation{The H. Niewodnicza\'nski Institute of Nuclear Physics, 
Polish Academy of Sciences, PL-31342 Krak\'ow, Poland}

\author{Jan Pluta}
\email{pluta@if.pw.edu.pl}
\affiliation{Faculty of Physics, Warsaw University of Technology, PL-00661 Warsaw, Poland}

\date{9 February 2006}

\begin{abstract}
Effects of the choice of the freeze-out hypersurface and resonance decays on the HBT interferometry in relativistic heavy-ion collisions are studied in detail within a class of models 
with single freeze-out. The Monte-Carlo method, as implemented in {\tt THERMINATOR}, is used to generate hadronic events describing production of particles from a thermalized and expanding source. All well-established hadronic resonances are included in the analysis as their role is crucial at large freeze-out temperatures. We use the two-particle method to extract the correlation functions, which allows us to study the Coulomb effects. We find that the pion HBT data from RHIC are fully compatible with the single freeze-out scenario, pointing at the shape of the freeze-out hypersurface where the transverse radius is decreasing with time. Results for the single-particle spectra for this situation are also presented. Finally, we present predictions for the kaon femtoscopy.
\end{abstract}



\maketitle 

\section{Introduction}

Femtoscopy is one of the most important and promising techniques used in relativistic heavy-ion collisions, as it reveals the spacetime characteristics of the fireball formed in the reaction. The study of two-particle correlations of identical particles is known as the Hanbury-Brown-Twiss (HBT) interferometry \cite{Baym:1997ce,Wiedemann:1999qn,Heinz:1999rw,Weiner:1999th,Tomasik:2002rx}, 
while together with the extension to non-identical particles it has been generically termed {\em femtoscopy} \cite{Lednicky:1990pu,Lednicky:2002fq}, referring to studies of the system at the femtometer scale. For a recent review of various aspects, both theoretical and experimental, of the field the reader is referred to the review by Lisa, Pratt, Soltz, and Wiedemann \cite{Lisa:2005dd}. There are several questions concerning the heavy-ion data for the HBT radii. The major puzzle is the practically constant  value of the HBT radii over the huge reaction energy range $\sqrt{s_{NN}}=20-200~{\rm GeV}$. The other surprising, when confronted to expectations based on earlier model predictions, feature found at RHIC is the proximity of the value of $R_{\rm out}/R_{\rm side}$ to unity, indicating a very short emission time of pions from the source. These challenges are to be faced by theoretical descriptions. As a matter of fact, a simultaneous reproduction of all HBT radii poses a serious problem to models with limited parametric freedom, as well as to hydrodynamical simulations or transport codes (for a recent study see \cite{Li:2006eh}, where $R_S$ and $R_L$ are found to be in agreement with data, while $R_O$ is predicted to be larger that the observed one). The above puzzles and problems led to significant revision of our understanding of the RHIC physics, with such important conclusions as the absence of the latent heat in hadronization \cite{Lisa:2005dd}, which would lead to a much longer-lived fireball, or the introduction of the concept of the length of homogeneity \cite{Makhlin:1987gm}, which effectively reduces the observed radii to smaller values than the geometric size of the whole source. 

In the present paper we study the femtoscopy at RHIC in a class of hydro-inspired models,
all with a single-freeze-out \cite{Broniowski:2001we}. Our analysis uses {\tt THERMINATOR }\cite{Kisiel:2005hn} - the THERMal heavy IoN generATOR for the single-freeze-out approach, which is a very flexible tool for studies of this type. The single freeze-out concept, which identifies the thermal and kinetic freeze-outs, may be viewed as an approximation to a more detailed evolution, taking into account different time scales for various processes. Nevertheless, the single freeze-out complies to the {\em explosive scenario} at RHIC \cite{Rafelski:2000by} and is definitely worth a detailed study in the context of femtoscopy. Moreover, the approach reproduces very efficiently the particle abundances, the transverse-momentum spectra, including particles with strangeness \cite{Broniowski:2001uk}, produces very reasonable results for the resonance production 
\cite{Broniowski:2003ax}, the balance functions in rapidity \cite{Bozek:2003qi}, the elliptic flow \cite{Broniowski:2002wp}, and the transverse energy \cite{Prorok:2004wi}. Approximate predictions of the model for the HBT radii were already presented in Ref.~\cite{Broniowski:2002wp,Broniowski:2002nf}. 

In particular, we deal in detail with two important issues. The first one is the influence of the chosen freeze-out hypersurface $\Sigma$ on the model prediction for the pionic HBT radii. We study several parametrization of $\Sigma$: the one from the original single-freeze-out model \cite{Broniowski:2001we}, as well as three hypersurfaces from the Blast-Wave model \cite{Schnedermann:1993ws,Retiere:2003kf} and its extension which account for the changes of the shape in the $t - \rho$ (time - transverse radius) space with the help of a single parameter $a$. We find significant dependence here, especially for the $R_{\rm long}$ radius. The analysis favors the shape of the freeze-out hypersurface where the transverse radius is decreasing with time. We also present the results for the single-particle spectra for the best freeze-out model out of the four studied, which turns out to be the modified Blast-Wave model with parameter $a=-0.5$. Clearly, the HBT correlation data place constraints on the freeze-out geometry and our study helps to put quantitative bounds on model parameters.

The second issue studied is the detailed role of hadronic resonances on the two-particle pion correlation. Although the basic features have been known since the very early days of the field \cite{Baym:1997ce,Akkelin:2004he}, there have been no HBT studies at large temperatures ($T \sim 170$~MeV) taking into account sufficiently many resonances. The inclusion of practically all resonances at such temperatures is important, as revealed by the studies of particle ratios \cite{Braun-Munzinger:2001ip,Rafelski:2004dp,Yen:1998pa,Gazdzicki:1998vd,Cleymans:2004pp,Becattini:2003wp} and the transverse-momentum spectra \cite{Broniowski:2001we}, where the resonances significantly decrease the inverse-slope parameter \cite{Florkowski:2001fp}. The presence of resonances plays an essential role also in femtoscopy supplying the separation distributions with long exponential tails, thus providing non-gaussian features in the HBT correlation functions. For a very recent study see Ref. \cite{Frodermann:2006sp}.  We should stress, that the models termed here as ``Blast-Wave'' have the blast-wave geometry, however they {\em do include all resonances}, contrary to many applications, where no resonance feeding is present. 

Finally, we present predictions for the {\em kaon} femtoscopy. We find that when the pion and kaon results are plotted together, they comply to the $m_T$-scaling conjuncture.

The paper is constructed as follows: In Sect. II we define a class of the hydro-inspired models with resonances which are used in our analysis of the correlation functions. Section III is a short introduction to the HBT formalism. In Sect. IV we present our main results on the HBT radii. The effects of the resonances on the correlation function are discussed in Sect. V. Section VII contains our predictions concerning the kaon correlation functions. 

\begin{figure}[t]
\begin{center}
\includegraphics[angle=0,width=0.47\textwidth]{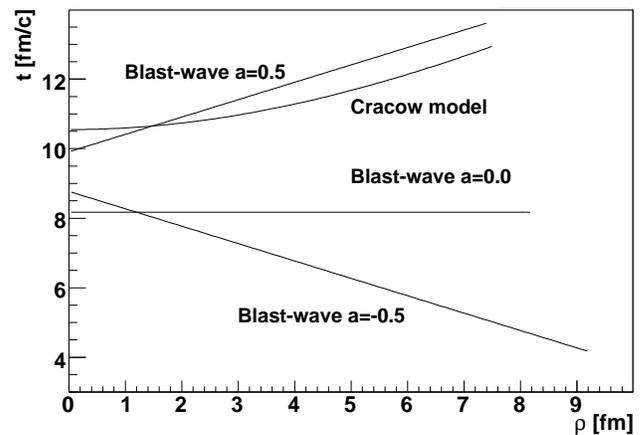}
\end{center}
\caption{Various parametrization of the freeze-out hypersurface. The curves show the dependence of time $t$ on the radial distance $\rho=\sqrt{r_x^2+r_y^2}$ at $r_z=0$
for the four models considered.}
\label{fig:models}
\end{figure}

\section{Hydro-inspired models of freeze-out}

The hydro-inspired models have become a very popular tool to analyze the data collected in relativistic heavy-ion collisions \cite{Florkowski:2004tn,Florkowski:2005nh}. The most popular 
model belonging to this class is the Blast-Wave model of Schnedermann, Sollfrank and Heinz 
\cite{Schnedermann:1993ws}. In the original formulation, it was designed to describe boost-invariant and cylindrically symmetric systems, hence it is best suited for description of the midrapidity region of the central Au + Au collisions studied at the top RHIC energies. 

Other models of this type include the Buda-Lund model \cite{Csorgo:1995bi,Csorgo:1999sj} and the Cracow single-freeze-out model \cite{Broniowski:2001we,Broniowski:2001uk,Broniowski:2002nf}. A distinctive feature of the Cracow model is that it includes the complete set of hadronic resonances. Precisely this feature allowed for the uniform description of the chemical and thermal freeze-outs.

In this work we consider the boost-invariant and cylindrically symmetric systems and use {\tt THERMINATOR} \cite{Kisiel:2005hn} to include the resonance effects. In this way, the Blast-Wave and Cracow model are treated on the same footing; {\em the contributions from the resonance decays that are very often neglected in the studies based on the Blast-Wave model are now taken completely into account}. The only (important) difference between the considered models resides in the definition of the freeze-out hypersurface; for the Blast-Wave model the freeze-out hypersurface is typically defined by the condition of the constant laboratory time, whereas for the Cracow model the freeze-out hypersurface is defined by the condition of the constant proper time. In the present paper we take into consideration these two cases and other possible options, all illustrated in Fig.~\ref{fig:models}.
The lines show the freeze-out hypersurfaces at  $r_{z} = 0$. 
Due to the measure $2\pi \rho d\rho$ the parts of the hypersurface with larger values of $\rho$ are more relevant in the evaluation of observables. 

\begin{figure}[t]
\begin{center}
\includegraphics[angle=0,width=0.43\textwidth]{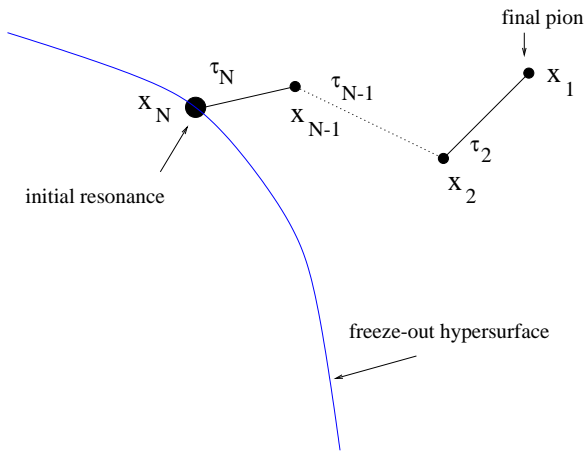}
\end{center}
\caption{The cascade of resonance decays: the initial resonance formed on the freeze-out hypersurface 
at the point $x_N$ with momentum $p_N$ decays after proper time 
$\tau_N$ at the point $x_{N-1}$. We track one decay product, which decays in sequence until the final 
pion is formed at the point $x_1$. }
\label{fig:decay}
\end{figure}

The boost-invariant, cylindrically symmetric freeze-out models are distinguished by different freeze-out curves in the $t-\rho$ space ($t$ is the laboratory time measured at $r_z = 0$, while $\rho=\sqrt{r_x^2+r_y^2}$ is the distance from the collision axis). The most popular Blast-Wave parametrization uses the freeze-out condition $t =$ const. In our calculations we also consider parametrizations of the form $t = \tau + a \, \rho$, where $\tau$ and $a$ are constants. For $a=0$ we reproduce the standard case. By studying the cases with different values of $a$ we may analyze the effect of the shape of the freeze-out hypersurface on different physical observables, including the HBT radii. The positive (negative) values of the parameter $a$ correspond to the freeze-out curves which go up (down) in the Minkowski $t-\rho$ space, {\em cf.} Fig.~\ref{fig:models}. It is important to realize that the freeze-out curves with negative $a$'s resemble the freeze-out curves obtained typically in hydrodynamic calculations. In the latter case the matter placed close to the surface of the system decouples earlier, while the matter inside the  system decouples later, only when the temperature in that region drops down due the expansion of the system. Consequently, by studying the effect of the choice of the freeze-out curve on the extracted HBT radii, we may find which freeze-out models are favored by the data.

\subsection{Emission function}

Our starting point is the formula for the pion emission function which takes into account sequential decays of the resonances \cite{Bolz:1992hc,Broniowski:2001uk,Broniowski:2002nf}. A contribution from one particular decay chain $c$, illustrated in Fig. \ref{fig:decay}, is given by the following equation
\begin{widetext}
\begin{eqnarray}
&&S_{c}\left( x_{1},p_{1}\right)  =  E_{p_1} {dN \over d^3p_1 d^4x_1}
= %\nonumber \\ && 
\int \frac{d^{3}p_{2}}{E_{p_{2}}}
B\left( p_{2},p_{1}\right) \int d\tau _{2}\Gamma _{2}e^{-\Gamma _{2}\tau_{2}} 
\int d^{4}x_{2}\delta ^{\left( 4\right) }
\left( x_{2}+\frac{p_{2}\tau_{2}}{m_{2}}-x_{1}\right) \times \dots \nonumber \\
&& \times \int \frac{d^{3}p_{N-1}}{E_{p_{N-1}}}B\left( p_{N-1},p_{N-2}\right)
\int d\tau _{N-1}\Gamma _{N-1}e^{-\Gamma _{N-1}\tau _{N-1}} 
\int d^{4}x_{N-1}\delta ^{\left( 4\right) }
\left( x_{N-1}+\frac{p_{N-1}\tau _{N-1}}{m_{N-1}}-x_{N-2}\right) \nonumber  \\
&&\times \int \frac{d^{3}p_{N}}{E_{p_{N}}} B\left( p_{N},p_{N-1}\right) \int
d\tau _{N}\Gamma _{N}e^{-\Gamma _{N}\tau _{N} }
\int d\Sigma _{\mu } \left(
x_{N}\right) \,p_{N}^{\mu }\,\,\delta ^{\left( 4\right) }\left( x_{N}+\frac{%
p_{N}\,\tau _{N}}{m_{N}}-x_{N-1}\right) f_{N}\left[ p_{N}\cdot u\left(
x_{N}\right) \right]. 
\label{emissionfct}
\end{eqnarray}
\end{widetext}
Here $d\Sigma _{\mu }$  is a 3-dimensional element of the freeze-out hypersurface $\Sigma$, the position of a resonance that decouples on the freeze-out hypersurface $\Sigma$  is denoted by $x_N$, its four-momentum by $p^\mu_N$, and its mass by $m_N$. 
The function $f_N$ is the thermal distribution function depending on the product of the resonance four-momentum $p_N$ and the local four-velocity $u(x_N)$. The resonance decays at the spacetime point $x_{N-1}$ and produces the tracked
daughter particle with the four-momentum $p^\mu_{N-1}$. In the first step of the simulation of the decay, the proper time $\tau _{N}$ is generated randomly according to the exponential decay law $({1/\Gamma_N}) \exp(-\Gamma _{N} \tau _{N} )$, where $\Gamma _{N}$ is the resonance width. Then, in the second step, the position  $x_{N-1}$ is obtained from the formula $x^\mu_{N-1} = x^\mu_N + (p^\mu_N/m_N) \tau_N $. The momentum of the daughter particle $p^\mu_{N-1}$ is determined purely by the available phase space, as described in Refs. \cite{Broniowski:2002nf,Kisiel:2005hn}. In Eq. (\ref{emissionfct}) the phase space effects are taken into account in terms of the splitting functions $B\left( p_{N},p_{N-1}\right)$ defined in \cite{Broniowski:2002nf}. If the daughter particle is a resonance, the above scheme is repeated until the final stable particle (pion, kaon, nucleon or antinucleon) is produced. The spacetime position and four-momentum of the final particle is denoted as $x_1$ and $p_1$. The cascade character of the process is reflected by the iterative structure of Eq.~(\ref{emissionfct}). The complete emission function is obtained as the sum over all possible decay channels $c$,
\begin{equation}
S(x,p) = \sum_{c} S_c(x,p).
\label{es}
\end{equation}
We note that all considered particles are on the mass shell with the energies $E_{p}
= \sqrt{m^2 + p^2}$.

\subsection{Distributions of the primordial particles}

The Monte-Carlo simulation of the decay process starts with the generation of hadronic distributions at the moment of freeze-out. We shall refer to such distributions as the {\em primordial distributions}. All hadronic states appearing in the Particle Data Tables \cite{Hagiwara:2002fs} are included in this procedure. Since we consider a boost-invariant and cylindrically symmetric system, the primordial distributions in rapidity, $y$, and 
transverse momentum, $p_\perp$, are of the form 
\begin{eqnarray}
&& {dN \over dy d^2p_\perp} =  \int d\Sigma_\mu(x_N) p^\mu_N f \left(p_N \cdot u(x_N)\right)
\nonumber \\
&& = {1 \over (2\pi)^3}
\int\limits_0^{2\pi} d\phi 
\int\limits_{-\infty}^{\infty} d\alpha_\parallel \int\limits_0^1 d\zeta
\,\, \rho(\zeta) {\tilde \tau}(\zeta) \nonumber \\
&& \times \left[m_\perp 
\hbox{cosh}(\alpha_\parallel-y) {d\rho \over d\zeta} 
- p_\perp \cos(\phi-\varphi) {d{\tilde \tau} \over d\zeta} \right] \nonumber \\
& & \times \left\{  \vphantom{{d{\tilde \tau} \over d\zeta} }
\exp\left[\beta m_\perp \hbox{cosh}(\alpha_\perp(\zeta)) 
\hbox{cosh}(\alpha_\parallel-y)  \right. \right. \nonumber \\
& & \left. \left.  - \beta p_\perp  
\hbox{sinh}(\alpha_\perp(\zeta)) \cos(\phi-\varphi) - \beta \mu \right] 
\pm 1  \vphantom{{d{\tilde \tau} \over d\zeta} }
\right\}^{-1}.
\label{dNdydpt}
\end{eqnarray}
Here $m_\perp = \sqrt{m^2+p^2_\perp}$ is the transverse mass, and $\varphi$ is the momentum azimuthal angle. The quantities $\phi$, $\alpha_\parallel$, and $\zeta$ are used to parametrize the freeze-out hypersurface: $\phi$ is the azimuthal angle, $\tan\phi = r_y/r_x$, $\alpha_\parallel$ is the spacetime rapidity, $\alpha_\parallel = 1/2 \ln[(t+r_z)/(t-r_z)]$, while $\zeta$ parametrizes the freeze-out curve obtained 
as the projection of the freeze-out hypersurface on the $r_z=0$ plane. The freeze-out curve is defined by the mapping $\zeta \to ({\tilde \tau}(\zeta),\rho(\zeta))$ relating the freeze-out time and position. At $r_z=0$ the variable  ${\tilde \tau} = \sqrt{t^2-r_z^2}$ coincides with the laboratory time, whereas $\rho = \sqrt{r_x^2+r_y^2}$ is the distance from the collision axis. The function $\alpha_\perp(\zeta)$ conveniently parametrizes the transverse collective flow, $v_\perp(\zeta)=\hbox{tanh}\,\alpha_\perp(\zeta)$. We recall that the longitudinal flow has the form $v_z=r_z/t$ \cite{Bjorken:1983qr}. The variable $\beta$ appearing in Eq. (\ref{dNdydpt}) is the inverse temperature, $\beta=1/T$, whereas the plus-minus sign is related to the statistics (in Eq. (\ref{dNdydpt}) and in the expressions below the upper sign corresponds to fermions, while the lower sign to bosons). With the help of the modified Bessel functions, Eq.~(\ref{dNdydpt}) may be rewritten in the more compact form 
\begin{eqnarray}
&& {dN \over dy d^2p_\perp} = {1 \over 2\pi^2} \sum\limits_{n=1}^\infty (\mp)^{n+1}
e^{n \beta \mu}
\int\limits_0^1 d\zeta \,\, \rho(\zeta) {\tilde \tau}(\zeta) 
\nonumber \\ 
&& \times \left\{
m_\perp {d\rho \over d\zeta} 
K_1 \left[n \beta m_\perp \hbox{cosh}\alpha_\perp \right]
I_0 \left[n \beta p_\perp \hbox{sinh}\alpha_\perp \right]  \right. \nonumber \\
&& \left. - \, p_\perp { d{\tilde \tau} \over d\zeta} 
K_0 \left[n \beta m_\perp \hbox{cosh}\alpha_\perp \right]
I_1 \left[n \beta p_\perp \hbox{sinh}\alpha_\perp \right] \right\}.
\label{dNdydptKI}
\end{eqnarray}

\subsection{Cracow single-freeze-out model}

Different boost-invariant and cylindrically symmetric models may differ in the
choice of the freeze-out curve  $({\tilde \tau}(\zeta),\rho(\zeta) )$. In the Cracow single-freeze-out model \cite{Broniowski:2001we,Broniowski:2001uk,Broniowski:2002nf} the freeze-out hypersurface is specified by the following equations:
\begin{equation}
{\tilde \tau} = \tau \, \hbox{cosh}\,\alpha_\perp(\zeta), \quad
\rho = \tau \, \hbox{sinh}\,\alpha_\perp(\zeta), \quad
\tau = \hbox{const}. \,\, ,
\label{crapar}
\end{equation}
which are equivalent to the condition
\begin{equation}
{\tilde \tau}^2 - \rho^2 = t^2 - r_z^2 -r_x^2 -r_y^2 = \tau^2.
\label{crainvtime}
\end{equation}
The parameterization (\ref{crapar}) implies that the freeze-out of the fluid elements placed farther away from the center happens at later times, see Fig. \ref{fig:models}. The velocity profile in the Cracow model has the Hubble-like structure
\begin{equation}
\vec{v}=\frac{\vec{r}}{t}. 
\label{velo}
\end{equation}
The use of Eqs. (\ref{crapar}) and (\ref{velo}) in Eq. (\ref{dNdydpt}) and the change of the integration variable $\zeta$ (first to $\alpha_\perp$ and later to $\rho$) leads to the distribution of the primordial particles in the 6-dimensional space of spacetime positions and momenta
\begin{widetext}
\begin{eqnarray}
{dN \over dy d\varphi p_\perp dp_\perp d\alpha_\parallel d\phi \rho \, d\rho } 
&=& {1 \over (2\pi)^3}
\left[m_\perp \sqrt{\tau^2+\rho^2} \, \hbox{cosh}(\alpha_\parallel-y) 
- p_\perp \rho \cos(\phi-\varphi) 
\right] \nonumber \\
& & \times \left\{
\exp\left[\beta m_\perp  \, \sqrt{1+ {\rho^2 \over \tau^2}} 
\hbox{cosh}(\alpha_\parallel-y) - \beta p_\perp  
{\rho \over \tau} \cos(\phi-\varphi) - \beta \mu \right] 
\pm 1 \right\}^{-1}.
\label{modA3}
\end{eqnarray}
\end{widetext}

\subsection{Generalized Blast-Wave Model}

As the second option considered in our calculations we choose the generalized Blast-Wave parameterization
\begin{equation}
{\tilde \tau} = \tau + a \rho, \quad
\hbox{tanh}\,\alpha_\perp(\zeta) = v_\perp = \hbox{const}.
\label{par2}
\end{equation}
The parameters: $\tau, a$ and $v_\perp$ are constants (for simplicity we assume that the transverse flow profile is constant). For $a=0$ we obtain the standard Blast-Wave parameterization corresponding to the assumption that the freeze-out process happens at constant laboratory time $t = \tau$ (in the central region where $\alpha_\parallel=0$). For $a > 0$ ($a < 0$ ) the straight line defining the freeze-out in the Minkowski space at $r_z=0$ goes upwards (downwards). Similarly to the previous case, the use of the 
parameterization (\ref{par2}) in Eq. (\ref{dNdydpt}) gives the 6-dimensional density
\begin{widetext}
\begin{eqnarray}
{dN \over dy d\varphi p_\perp dp_\perp d\alpha_\parallel  d\phi \rho  d\rho \, } 
&=& {1 \over (2\pi)^3} (\tau + a\rho)
\left[ m_\perp \hbox{cosh}(\alpha_\parallel-y) 
- a \, p_\perp \cos(\phi-\varphi)\right] \nonumber \\
& & \times \left\{
\exp\left[ { \beta m_\perp  \over \sqrt{1 - v_\perp^2}} \,\,
\hbox{cosh}(\alpha_\parallel-y) - { \beta p_\perp   v_\perp \over \sqrt{1 - v_\perp^2}}
 \cos(\phi-\varphi) - \beta \mu \right] 
\pm 1 \right\}^{-1}. 
\label{modA2}
\end{eqnarray}
\end{widetext}

%%%%%%%%%%%%%%%%%%%%%%%%%%%%%%%%%%%%%%%%%%%%%%%%%%%%%%%%%%%%%%%%%%%%%%%%%%%%%%%%%%
%%%%%%%%%%%%%%%%%%%%%%%%%%%%%%%%%%%%%%%%%%%%%%%%%%%%%%%%%%%%%%%%%%%%%%%%%%%%%%%%%%
%%%%%%%%%%%%%%%%%%%%%%%%%%%%%%%%%%%%%%%%%%%%%%%%%%%%%%%%%%%%%%%%%%%%%%%%%%%%%%%%%%
\section{The HBT formalism}
\label{hbtformal}

%%%%%%%%%%%%%%%%%%%%%%%%%%%%%%%%%%%%%%%%%%%%%%%%%%%%%%%%%%%%%%%%%%%%%%%%%%%%%%%%%%
%%%%%%%%%%%%%%%%%%%%%%%%%%%%%%%%%%%%%%%%%%%%%%%%%%%%%%%%%%%%%%%%%%%%%%%%%%%%%%%%%%
\subsection{Two-particle correlation function}
\label{twopartcf}

Consider the two-particle distribution expressed by the two-particle emission function,
\begin{eqnarray}
W_2({\vec p}_1, {\vec p}_2) &=& E_{p_1} E_{p_2} \frac {dN} {d^3 p_1 d^3 p_2} \nonumber \\
&=& \int S(x_1, x_2, p_1, p_2) d^4 x_1 d^4 x_2.
\label{doublespectra}
\end{eqnarray}
The correlation function is then defined as
\begin{equation}
C({\vec p}_1, {\vec p}_2) = \frac {W_2({\vec p}_1, {\vec p}_2)} {W_1({\vec p}_1)
W_1({\vec p}_2)}
\label{cfbyemission}
\end{equation}
where
\begin{equation}
W_1({\vec p}) = E_p {dN \over d^3p} = \int d^4x\, S(x,p)
\label{spectrum}
\end{equation}
with $S(x,p)$ given by Eqs. (\ref{emissionfct}) and (\ref{es}).

One may assume that the two-particle production probability is influenced only by the two-particle interaction. In this case one neglects the many-body interactions between the produced particles as well as the event-wide correlations (e.g., the effects induced by the momentum conservation). Then, the two-particle emission function may be expressed as the product of the single-particle emission functions and the squared wave-function of the pair. After taking into account the {\it smoothness approximation} we write 
\begin{equation}
C({\vec q},{\vec k}) = \frac {\int d^4 x_1 S(x_1,p_1) d^4 x_2
S(x_2,p_2) | \Psi( {\vec k}^{*} , {\vec r}^{*}) |^2} { \int d^4 x_1
S(x_1, p_1) \int d^4 x_2 S(x_2, p_2)}.
\label{cfbypairwave}
\end{equation}
We define the momentum difference of the particles as
\begin{equation}
q = (q_0, {\vec q}) = \left(E_{p_1} - E_{p_2}, {\vec p}_1 - {\vec p}_2 \right), 
\label{vq}
\end{equation}
the sum of their momenta as 
\begin{equation}
P = (P_0, {\vec P}) = \left(E_{p_1} + E_{p_2}, {\vec p}_1 + {\vec p}_2 \right), 
\label{vP}
\end{equation}
and the average momentum of the pair as 
\begin{equation}
k = \left( k_0, {\vec k} \right) = {{1} \over {2}} 
\left( E_{p_1} + E_{p_2}, {\vec p}_1 + {\vec p}_2 \right).
\label{vK}
\end{equation}
The generalized momentum difference is defined by the formula
\begin{equation}
{\tilde q} = q - \frac{P(q \cdot P)}{P^2},
\label{tildeq1}
\end{equation}
which in the pair rest frame (PRF) is reduced to the form
\begin{equation}
{\tilde q} = \left(0, 2 \vec k^{*} \right).
\label{tildeq2}
\end{equation}
For the particles with equal masses we use the notation 
\begin{equation}
{\vec q}_{\rm inv} = 2 {\vec k}^*. 
\label{qinv}
\end{equation}
The space and time separations of the members of the pair are: $ {\vec r} = {\vec r}_{1} - {\vec  r}_{2}$ and $\Delta t = t_1 - t_2$. If calculated in PRF they are denoted as ${\vec r}^*$ and $\Delta t^*$.  Both ${\vec k}^*$ and ${\vec r}^*$ appear as the arguments of the wave function in Eq. (\ref{cfbypairwave}) since PRF is the most convenient reference frame for the representation of the wave function structure.  

In general, the HBT analysis may be performed in any reference frame. One determines the correlation function as a function of the relative-momentum components in the selected frame. Then the inverse widths of the correlation functions yield the size parameters of the system in this frame. In the present paper we use the Bertsch-Pratt decomposition \cite{Bertsch:1988db,Pratt:1986cc,Chapman:1994yv} of the average and relative three-momenta into three components. The {\it long} axis coincides with the beam axis, the {\it out} axis is determined by the direction of the average transverse momentum of the pair, denoted later by ${\vec k}_T$, and the {\it side} direction is perpendicular to the other two axes. Following the RHIC experiments we choose to perform the analysis in the longitudinal co-moving system (LCMS), which is defined as a system where $k_{\rm long}=0$. The HBT radii presented below are always obtained in the LCMS system. Later in this work the notation is used in which the values in PRF are denoted by an asterisk, while the values without asterisk are defined in LCMS.

By the definition of the Monte-Carlo method, the numerical equivalent of the integrals  (\ref{cfbypairwave}) is the summation over particles or pairs of particles generated by the Monte-Carlo procedure. The numerical calculation of the correlation functions is done in bins, which may be expressed with the help of the function
\begin{eqnarray}
\delta_{\Delta}({\bf x}) = 
\left\{
\begin{array}{cc}
1 & \hbox{if}  \,\,\,  | {\bf x} | \leq  \frac{\Delta}{2},  \\
& \\
0 & \hbox{otherwise}.
\end{array}
\right.
\label{deltadelta}
\end{eqnarray}     
Then the correlation function may be expressed simply as
\begin{eqnarray}
&& \!\!\!\!\!\!C({\vec q}, {\vec k}) = \nonumber \\
&& \!\!\!\!\!\!\frac{\sum\limits_{i} \sum\limits_{j \neq i} \delta_\Delta({\vec q} 
- {\vec p}_i + {\vec p}_j ) \delta_\Delta({\vec k} - \frac{1}{2}({\vec p}_i + {\vec p}_j) )
|\Psi({\vec k}^{*}, {\vec r}^{*}) |^2} 
{\sum\limits_i \sum\limits_j \delta_\Delta({\vec  q} - {\vec p}_i + {\vec p}_j ) 
\delta_\Delta({\vec  k} - \frac{1}{2}({\vec p}_i + {\vec p}_j ))}. \nonumber \\
\label{cfbysum}
\end{eqnarray}
In our numerical calculations we use $\Delta$ = 5 MeV.

%%%%%%%%%%%%%%%%%%%%%%%%%%%%%%%%%%%%%%%%%%%%%%%%%%%%%%%%%%%%%%%%%%%%%%%%%%%%%%%%%%
%%%%%%%%%%%%%%%%%%%%%%%%%%%%%%%%%%%%%%%%%%%%%%%%%%%%%%%%%%%%%%%%%%%%%%%%%%%%%%%%%%
\subsection{Wave function of the pion pair}

Various analyses of the HBT correlations use different approximations for the full pion wave function $\Psi$. In the non-interacting system or in the interacting but non-relativistic case the motion of the center of mass can be separated and one deals with the relative motion only. The simplest relative wave function ignores all dynamical interactions and has the form
\begin{equation}
\Psi^{Q} = \frac{1} {\sqrt{2}} (e^{i {\vec k}^* {\vec r}^*} 
+ e^{-i {\vec k}^* {\vec r}^*}),
\label{psiq}
\end{equation}
where symmetrization over the two identical particles has been performed. Therefore, 

\begin{equation}
|\Psi^{Q}|^2 = 1 + \cos\left(2 {\vec k}^* {\vec r}^*\right).
\label{psiqsq}
\end{equation}
Correlation functions calculated according to (\ref{cfbysum}) and (\ref{psiqsq}) represent the ideal Bose-Einstein correlation functions. They are also very useful in the model studies, because they can be calculated analytically for simple gaussian emission functions. 

In more realistic calculations, the Coulomb interaction of the charged pion pairs should be taken into account, which may be achieved by the use of the wave function
\begin{eqnarray}
\Psi^{QC} &=& e^{i \delta_c} \sqrt{A_c(\eta)} \frac {1} {\sqrt{2}} 
\left[ e^{-i \vec k^* \vec r^*} F(-i
\eta, 1, i \xi^{+}) \right. \nonumber \\
&& + \left. e^{i \vec k^* \vec r^*} F(-i \eta, 1, i \xi^{-})\right],
\label{psiqc}
\end{eqnarray}
where $\delta_c$ is the Coulomb phase shift, $A_c$ is the Coulomb penetration factor (sometimes called the Gamow factor), $\xi^{\pm} = k^* r^*  \pm \vec k^* \vec r^*  = k^* r^* (1 \pm \cos\theta^*)$, $\eta = (k^* a)^{-1}$ with $a$ being the Bohr radius of the pair, and $F$ is the confluent hypergeometric function. $\theta^*$ is the angle between $k^*$ and $r^*$. The correlation function obtained in this way can be compared directly to the correlation function obtained from the experiment. The complete wave function of the pion pair contains a contribution from the strong interaction as well. However these are small in the isospin $I=2$ channel and are neglected.

\begin{figure}[t]
\begin{center}
\includegraphics[angle=0,width=0.48 \textwidth]{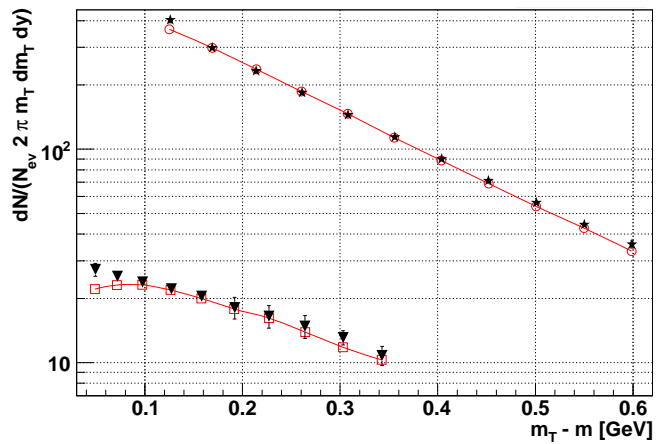}
\end{center}
\caption{Transverse-mass spectra at mid-rapidity of pions (open circles) and kaons (open squares) for the Blast-Wave model with $a=-0.5$, $T=165.6$ MeV, $\mu_B=28.5$ MeV, $\tau=8.55$ fm, $\rho_{\rm max}=8.92$ fm, and $v_{T} = 0.311$ c. The data points (stars for pions, triangles for kaons) come from the STAR collaboration \cite{Adams:2004yc}.  }
\label{fig:spectra}
\end{figure}

%%%%%%%%%%%%%%%%%%%%%%%%%%%%%%%%%%%%%%%%%%%%%%%%%%%%%%%%%%%%%%%%%%%%%%%%%%%%%%%%%%
%%%%%%%%%%%%%%%%%%%%%%%%%%%%%%%%%%%%%%%%%%%%%%%%%%%%%%%%%%%%%%%%%%%%%%%%%%%%%%%%%%

\subsection{Numerical calculation of correlation functions}

The correlation functions analyzed in this work are obtained through a numerical implementation of Eqs. (\ref{cfbysum}) and (\ref{psiq}) or Eqs. (\ref{cfbysum}) and (\ref{psiqc}). Particles generated by {\tt THERMINATOR} are grouped into events, as in experiment. In each event every charged pion is combined with every other pion of the same charge. For each pion pair, $|\Psi|^2$ is calculated and added to the numerator of Eq. (\ref{cfbysum}) in a bin corresponding to its $q_{\rm inv}$ for 1-dimensional functions or to its $q_{\rm out}$, $q_{\rm side}$ and $q_{\rm long}$ for the full 3-dimensional case. At the same time, 1 is added to the denominator of Eq. (\ref{cfbysum}) in the corresponding bin. The resulting ratio yields the correlation function. 

By making a proper selection of single pions and pairs of pions one may study the
correlation functions as functions of various variables. For instance, taking into account the pairs of particles within a certain total momentum range only, one immediately obtains the dependence on $k_T$. It is important to note that all single-particle approaches in the studies of the correlation functions use transverse-momenta of single particles only, while the experimental data are represented as functions of $k_T$. This is one of the advantages of the two-particle method over the single-particle method. Another one is the possibility of including final-state interactions, such as Coulomb effects.

\begin{figure}[t]
\begin{center}
\includegraphics[angle=0,width=0.42\textwidth]{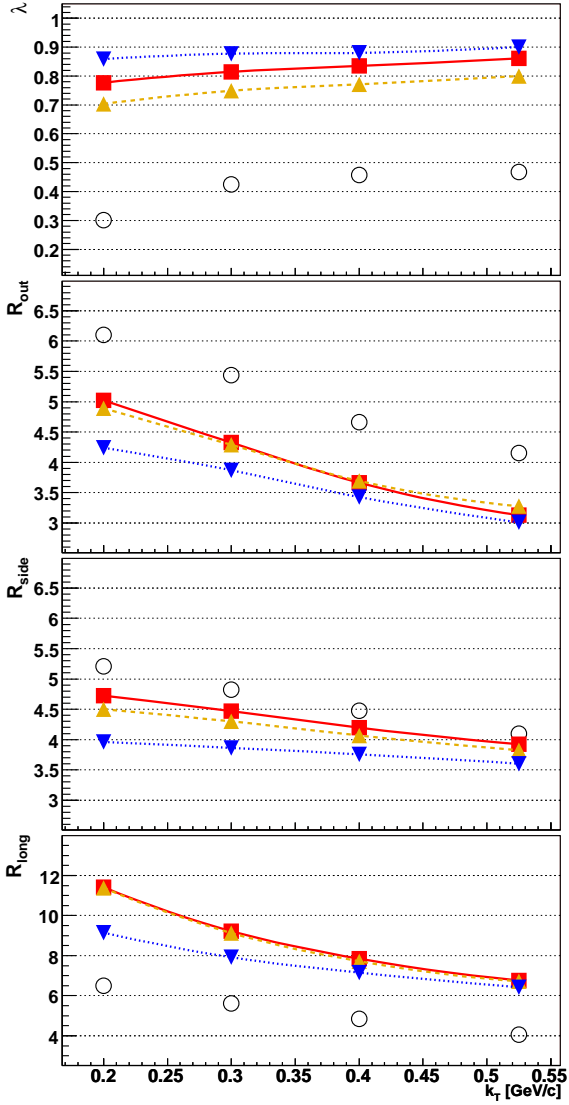}
\end{center}
\caption{The results for Cracow model: $\lambda$ and the HBT radii $R_{\rm out}$, $R_{\rm side}$, and $R_{\rm long}$ shown as functions of the transverse momentum of the pion pair. The squares show the full calculation with resonances based on the method of Sect. \ref{hbtformal}, down-triangles is the same without resonances, the up-triangles show the calculation with resonances and the Coulomb corrections made according to the Bowler-Sinyukov method, while the circles show the data of the STAR collaboration for 
$\sqrt{s_{NN}}=200$~GeV~\cite{Adams:2004yc}. The lines are drawn to guide the eye. We note that the inclusion of resonances increases the radii by about 1 fm. The model parameters are: $T=165.6$ MeV, $\mu_B=28.5$ MeV, $\tau=10.55$ fm, and $\rho_{\rm max}=7.53$ fm.}
\label{fig:cra}
\end{figure}

\begin{figure}[t]
\begin{center}
\includegraphics[angle=0,width=0.42\textwidth]{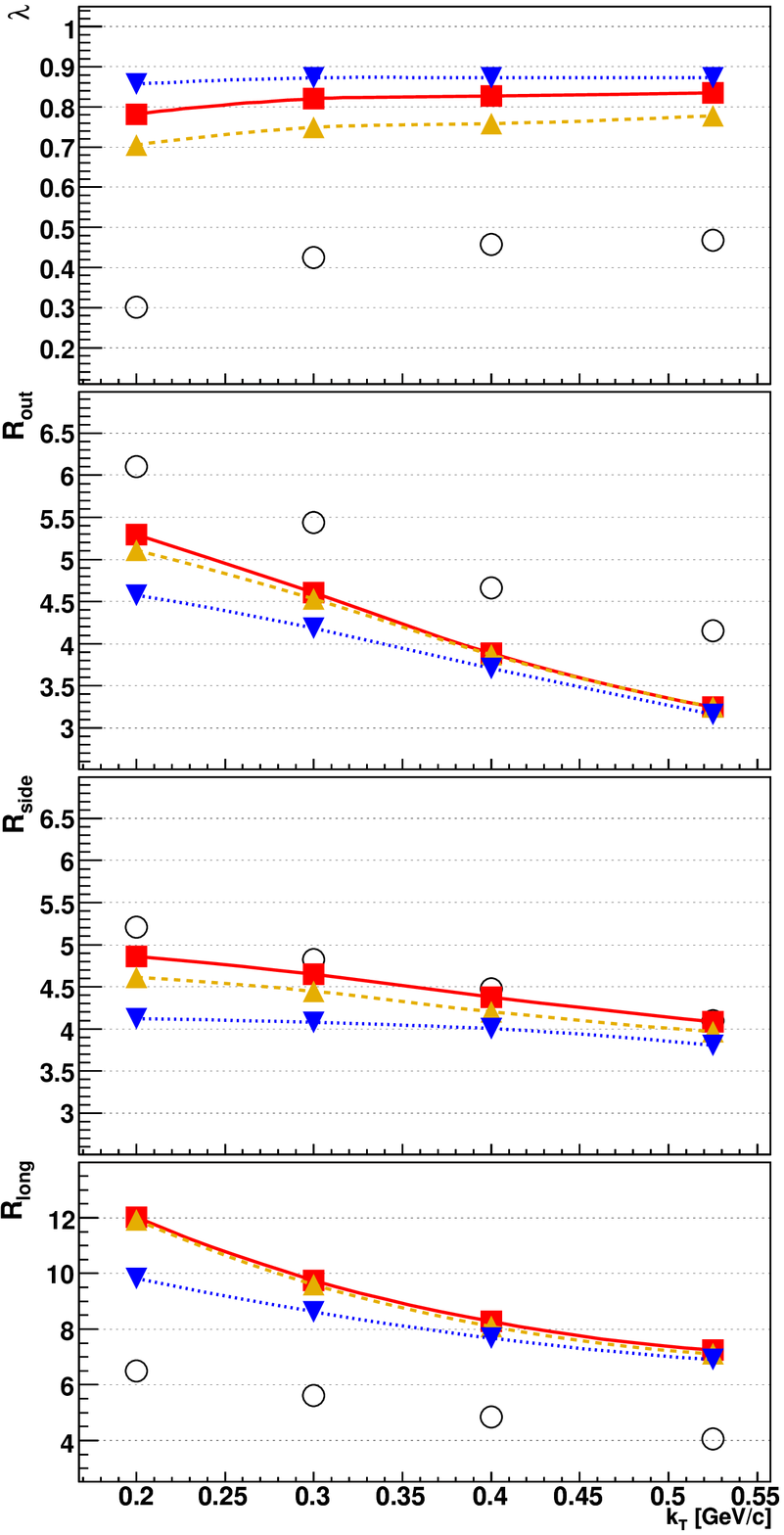}
\end{center}
\caption{Same as Fig. \ref{fig:cra} for the Blast-Wave model with resonances, $a=0.5$.
The model parameters are: $T=165.6$~MeV, $\mu_B=28.5$~MeV, $\tau=9.91$~fm,
$\rho_{\rm max}=7.43$~fm, and $v_{T}=0.407$~c.}
\label{fig:bwp}
\end{figure}

\begin{figure}[t]
\begin{center}
\includegraphics[angle=0,width=0.42\textwidth]{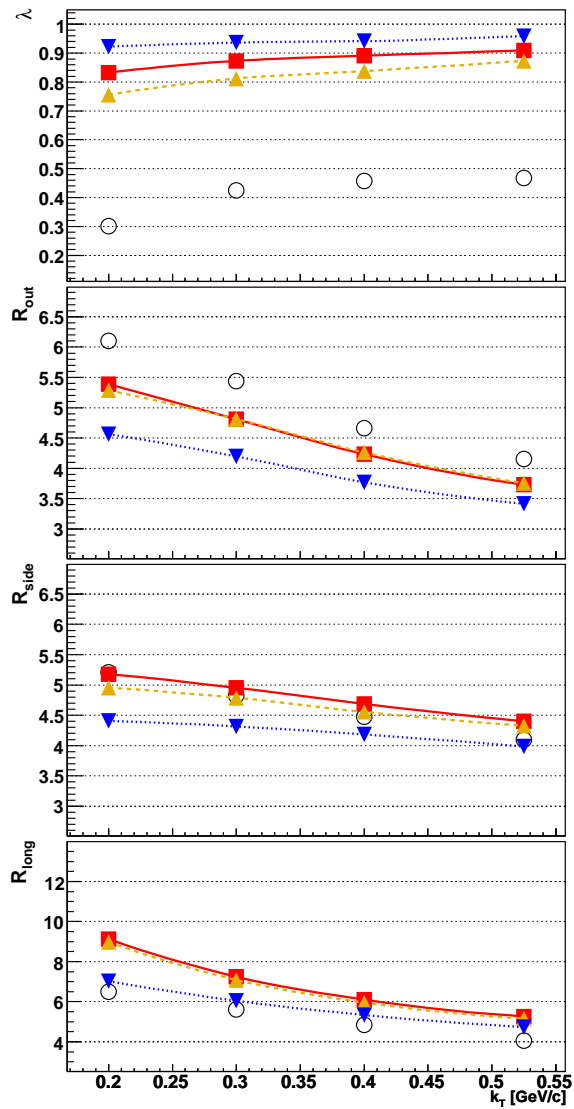}
\end{center}
\caption{Same as Fig. \ref{fig:cra} for the Blast-Wave model with resonances, $a=0$.
The model parameters are: $T=165.6$~MeV, $\mu_B=28.5$~MeV, $\tau=8.17$~fm,
$\rho_{\rm max}=8.21$~fm, and $v_{T}=0.341$~c.}
\label{fig:bw0}
\end{figure}

\begin{figure}[t]
\begin{center}
\includegraphics[angle=0,width=0.42\textwidth]{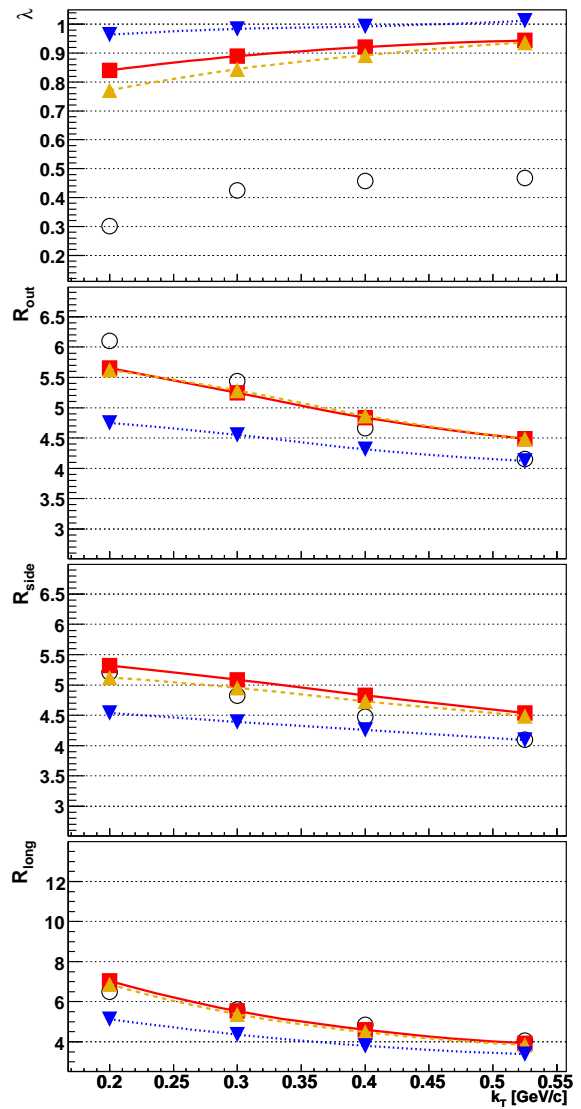}
\end{center}
\caption{Same as Fig. \ref{fig:cra} for the Blast-Wave model with resonances, $a=-0.5$.
The model parameters are: $T=165.6$~MeV, $\mu_B=28.5$~MeV, $\tau=8.55$~fm, $\rho_{\rm max}=8.92$~fm, and $v_{T}=0.311$~c. This is the model that produces best agreement out of the four models considered. With the same values of the parameters the model reproduces the transverse-mass spectra, see Fig. \ref{fig:spectra}.}
\label{fig:bwm}
\end{figure}

\begin{figure}[t]
\begin{center}
\includegraphics[angle=0,width=0.42\textwidth]{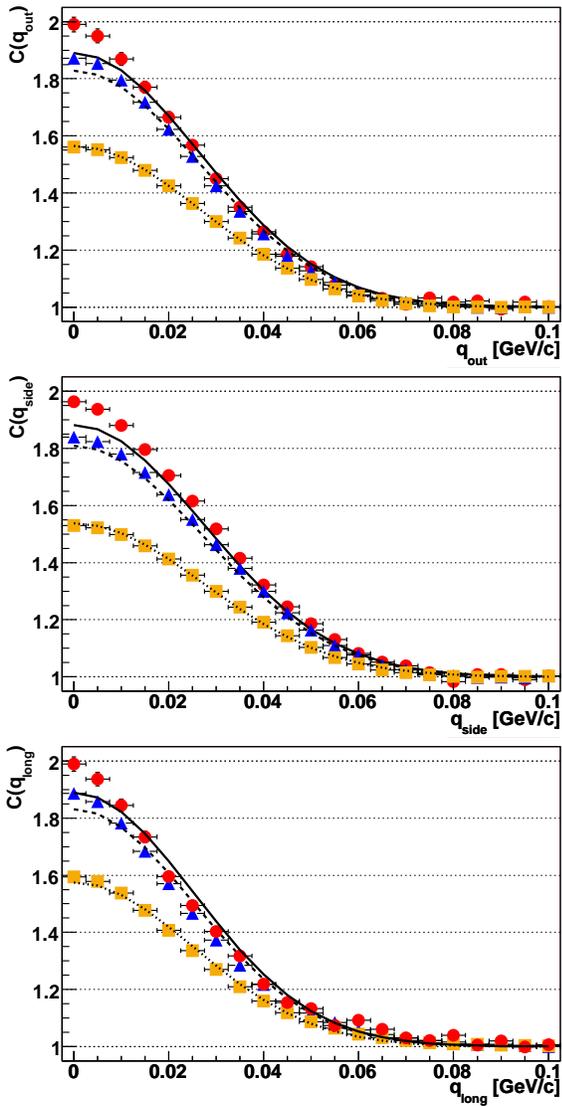}
\end{center}
\caption{An example of the projected pion correlation functions for the Blast-Wave model with resonances, $a=-0.5$. Model parameters are the same as for Fig. \ref{fig:bwm}. The correlation functions include pion pairs with transverse-momentum in the range: 0.25 GeV < $k_T$ < 0.35 GeV. We show the projections of the correlation function (symbols) and the projections of the 3-dimensional fit (lines). The top plot shows projections on the $q_{\rm out}$ axis, i.e., the 3-dimensional correlation function has been integrated in two other directions over some range. The middle plot shows the projection on $q_{\rm side}$, and the bottom plot the projection on $q_{\rm long}$. Circles (solid lines) show the function (fit) integrated in two other directions in the range: $|q_i|$ < 2.5 MeV, triangles (dashed lines) for $|q_i|$ < 12.5 MeV, and squares (dotted line) for $|q_i|$ < 32.5 MeV, where $i$ = {\it out}, {\it side} or {\it long}. }
\label{fig:cf3dqs}
\end{figure}

\begin{figure}[t]
\begin{center}
\includegraphics[angle=0,width=0.42\textwidth]{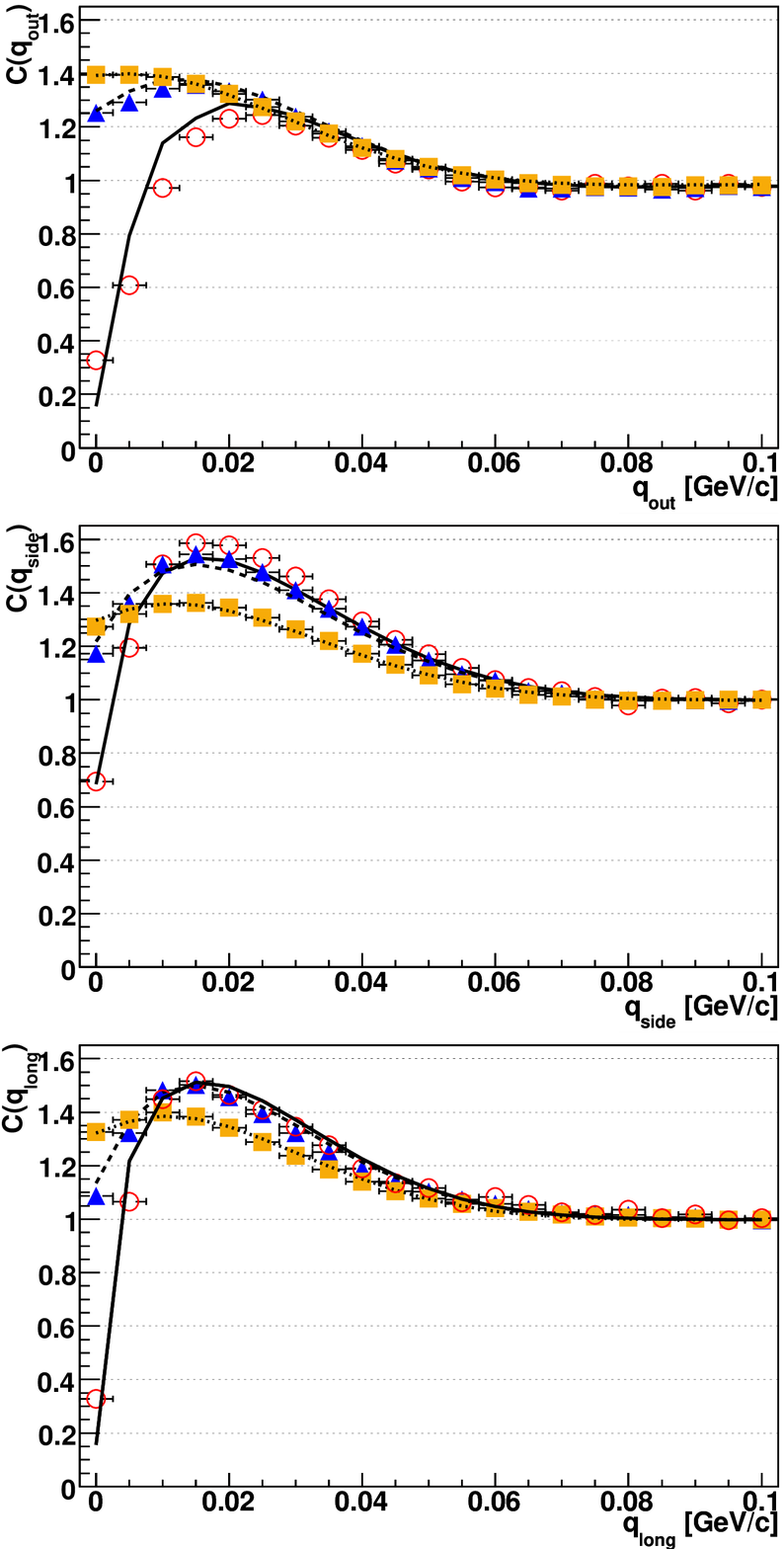}
\end{center}
\caption{The same as Fig. \ref{fig:cf3dqs} for the analysis including the Coulomb interaction.}
\label{fig:cf3dqsc}
\end{figure}

\subsection{Extraction of HBT radii}
\label{getrad}

In most of the realistic cases the integral (\ref{cfbypairwave}) cannot be performed analytically. One of the cases where the analytic calculation may be done corresponds to the situation where the pion wave function is given by Eq.~(\ref{psiq}) and the single-particle emission function is a static 3-dimensional ellipsoid with a gaussian density profile
\begin{eqnarray}
S(\vec x, \vec p) = N \exp\left(-{x_{\rm out} ^2 \over {2 R_{\rm out}^2}} -
{x_{\rm side}^2 \over {2 R_{\rm side}^2}} 
- {x_{\rm long}^2 \over {2 R_{\rm long}^2}} \right).
\label{staticsource}
\end{eqnarray}
Please note that this source function is static - it does not depend on particle momentum.  In this case the integral (\ref{cfbypairwave}) with the free wave function (\ref{psiq}) leads to the well known formula
\begin{eqnarray}
& & C\left(k_\perp,q_{\rm out},q_{\rm side},q_{\rm long} \right) = 1 + 
\lambda \exp\left[
-R^2_{\rm out}(k_\perp) q^2_{\rm out} \right.
\nonumber \\
& & \left.
-R^2_{\rm side}(k_\perp) q^2_{\rm side}
-R^2_{\rm long}(k_\perp) q^2_{\rm long}
\right].
\label{cfgaus}
\end{eqnarray}
The quantities $R_{\rm out}$, $R_{\rm side}$ and $R_{\rm long}$, known as the ``HBT radii'' are the widths of the gaussian approximation to the single-particle freeze-out distribution. It is important to emphasize that formula (\ref{cfgaus}) is commonly used to fit the experimental data and to represent the results of the model calculations although the experimental or model emission functions are frequently far from gaussians. In context of our model this issue will be discussed in detail in Sect. V.

%%%%%%%%%%%%%%%%%%%%%%%%%%%%%%%%%%%%%%%%%%%%%%%%%%%%%%%%%%%%%%%%%%%%%%%%%%%%%%%%%%
%%%%%%%%%%%%%%%%%%%%%%%%%%%%%%%%%%%%%%%%%%%%%%%%%%%%%%%%%%%%%%%%%%%%%%%%%%%%%%%%%%

\subsection{Bowler-Sinyukov formalism}

The pion correlation functions for sizes typically observed in heavy-ion collisions are significantly influenced by the Coulomb interaction, hence the model calculations should also include this effect. We take into account the Coulomb interaction by using the exact form of the two-particle wave-function (\ref{psiqc}). In the procedure of including the Coulomb effects we should accordingly modify the form of the reference function (\ref{cfgaus})  that is used to extract the HBT radii. Since there is no analytic formula which parametrizes the Coulomb effects exactly, we follow the procedure of CERES \cite{Adamova:2002wi}, STAR \cite{Adams:2004yc} and PHENIX \cite{Adler:2003kt}, which is based on the following assumptions: the Coulomb interaction and the wave-function symmetrization factorize and, moreover, the Coulomb interaction part of the function can be replaced by the averaged Coulomb wave-function. With these assumptions the Coulomb interaction may be separated from the integration in the numerator of (\ref{psiqc}) and one obtains the Bowler-Sinyukov formula \cite{Bowler:1991vx,Sinyukov:1998fc}
\begin{eqnarray}
&& C(\vec q, \vec k) = (1 - \lambda) + \lambda K_{\rm coul}(q_{\rm inv})
\left[1 +  \exp \left(-R_{\rm out}^2 q_{\rm out}^2  \right. \right.
\nonumber \\
&& \left. \left.
- R_{\rm side}^2 q_{\rm side}^2 - R_{\rm long}^2
q_{\rm long}^2 \right)  \right],
\label{cffitbs}
\end{eqnarray}
where $K_{\rm coul}(q_{\rm inv})$ is the squared Coulomb wave function integrated over a static gaussian source. We use, following the STAR procedure \cite{Adams:2004yc}, the static gaussian source characterized by the widths of 5 fm in all three directions. The 3-dimensional correlation function with the exact treatment of the Coulomb interaction,  calculated according to Eqs. (\ref{cfbypairwave}) and (\ref{psiqc}), is then fitted with this approximate formula and the HBT radii are obtained. They can be compared directly to the experimental radii. We note that this way of comparing experimental data and theoretical predictions is very reasonable in the sense that the same experimental and theoretical observables are compared. The HBT radii may be obtained also from the fit to the correlation function (\ref{cfgaus}), which provides the way for the experiments to judge the systematic uncertainty of determining the results by using the Bowler-Sinyukov procedure.

\begin{figure}[t]
\begin{center}
\includegraphics[angle=0,width=0.47\textwidth]{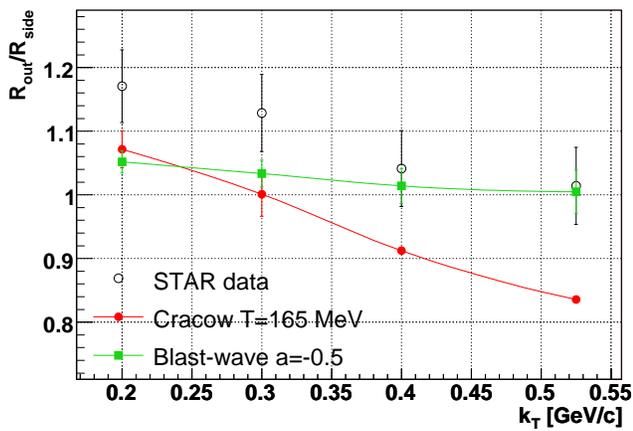}
\end{center}
\caption{Ratio $R_{\rm out}/R_{\rm side}$ for the Cracow model with $T=165$ MeV (circles) and the Blast-Wave model with resonances and $a=-0.5$ (squares). The STAR data are represented by open circles. }
\label{fig:rors}
\end{figure}

%%%%%%%%%%%%%%%%%%%%%%%%%%%%%%%%%%%%%%%%%%%%%%%%%%%%%%%%%%%%%%%%%%%%%%%%%%%%%%%%%%
%%%%%%%%%%%%%%%%%%%%%%%%%%%%%%%%%%%%%%%%%%%%%%%%%%%%%%%%%%%%%%%%%%%%%%%%%%%%%%%%%%
%%%%%%%%%%%%%%%%%%%%%%%%%%%%%%%%%%%%%%%%%%%%%%%%%%%%%%%%%%%%%%%%%%%%%%%%%%%%%%%%%%
\section{Results}

The parameters of each model are fixed by fitting the single-particle $p_\perp$-spectra of pions and kaons to the experimental data. An example of such a fit is shown in Fig.~\ref{fig:spectra}. The values of the parameters are listed in the captions of Figs.~\ref{fig:cra}-\ref{fig:bwm}. Thus our analysis of the HBT correlations has no extra parametric freedom.

Our basic results for the pion interferometry are shown in Figs.~\ref{fig:cra}-\ref{fig:bwm}. They were obtained by the procedure described in detail in Sect.~\ref{hbtformal}, i.e., by fitting the 3-dimensional two-particle correlation functions. In Figs.~\ref{fig:cra}-\ref{fig:bwm} we show the intercept $\lambda$ and the HBT radii $R_{\rm out}$, $R_{\rm side}$, and $R_{\rm long}$ as functions of the transverse momentum of the pion pair. The squares correspond to the full calculation with resonances, the down-triangles show the results obtained in the calculation without resonances, the up-triangles show the results obtained with resonances and with the Coulomb-aware fit made according to the Bowler-Sinyukov formalism \cite{Bowler:1991vx,Sinyukov:1998fc}, while the circles show the data of the STAR collaboration from Ref.~\cite{Adams:2004yc}. The first immediate observation is that the inclusion of resonances increases the radii by about 1 fm. This is expected, since the resonances travel some distance from their place of birth on the freeze-out hypersurface before they decay into pions. The typical scale is set by the resonance life-time which is about 1 fm/c. 

\begin{figure*}[t]
\begin{center}
\includegraphics[angle=0,width=0.8\textwidth]{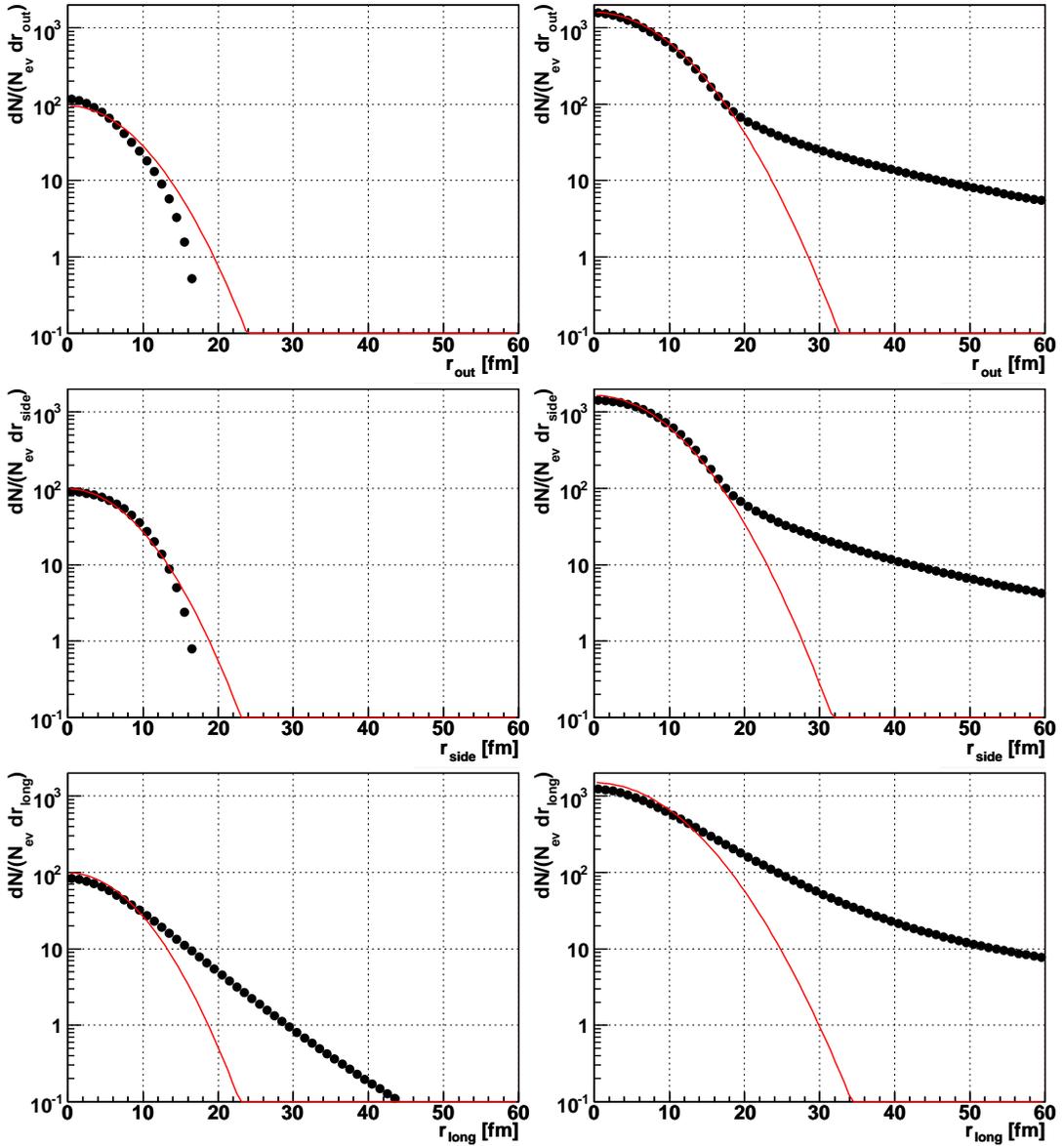}
\end{center}
\caption{The separation distributions for pion pairs from the 
Blast-Wave model with resonances and $a=-0.5$ (black circles). On the left-hand-side the
plots for pairs of primordial pions are shown. On the right-hand-side,
the plots for all pions are shown. The lines show the separation
distribution which is the result of the fitting of the corresponding
correlation function by a gaussian parameterization. 
}
\label{fig:res}
\end{figure*}

\begin{figure}[t!]
\begin{center}
\includegraphics[width=0.45\textwidth]{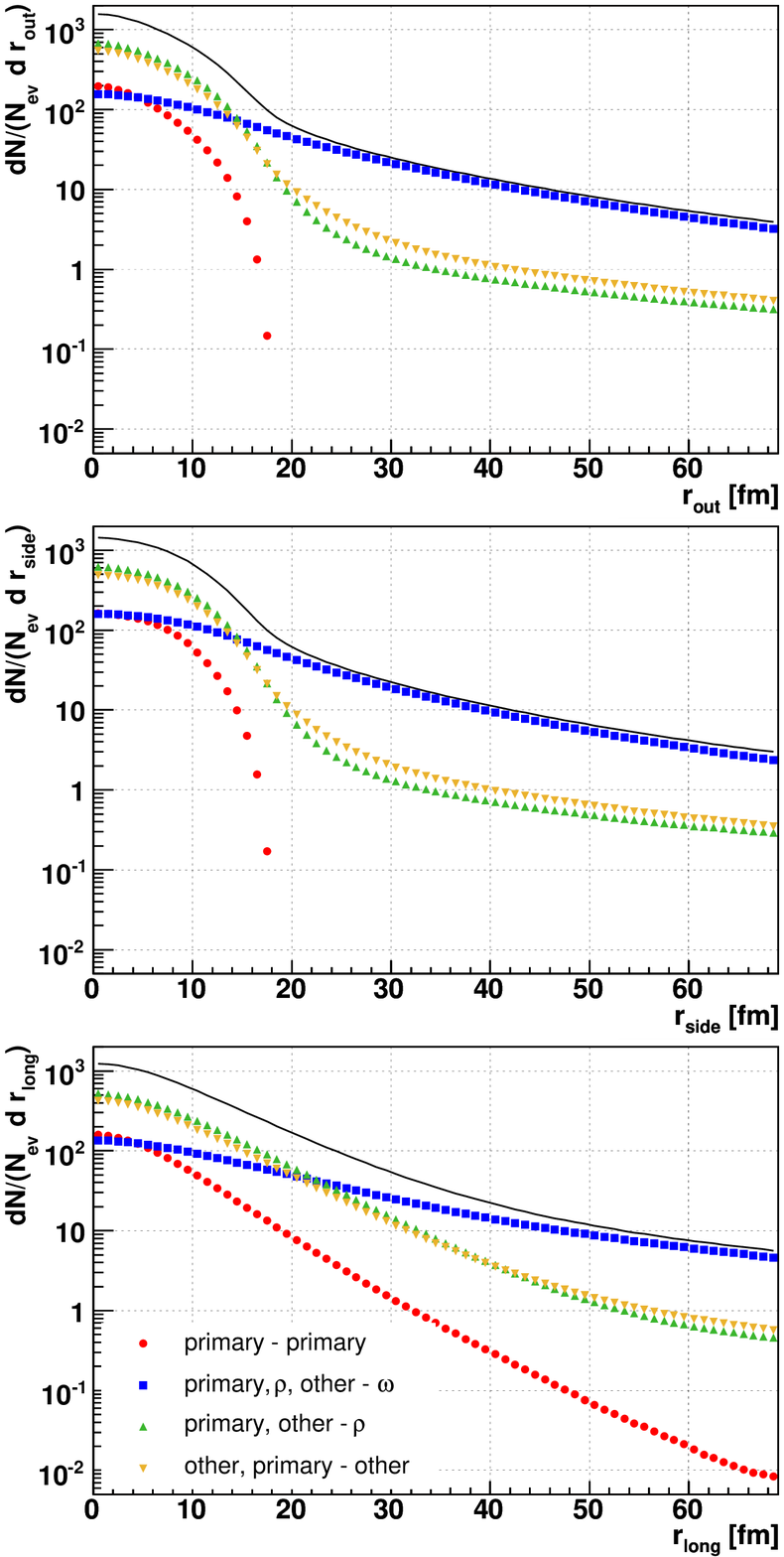}
\end{center}
\caption{Space distribution of the produced pion pairs. The solid lines show separation distribution for all pairs, circles show the pairs of primordial pions, up-triangles show the pairs which contain one pion from the $\rho$ decay, but no pion from the $\omega$ decay, squares show the pairs where one pion comes from the $\omega$ decay, finally down-triangles show the pairs where one pion comes from decays of resonances other than $\rho$ and $\omega$ and the other from this group plus the primordial pions, see text for a more detailed description. Blast-Wave model with resonances, $a=-0.5$, 0.25 GeV < $k_{T}$ < 0.35 GeV.}
\label{fig:anatom}
\end{figure}

Browsing through Figs.~\ref{fig:cra}-\ref{fig:bwm} we note that the agreement with the data changes with the selected model of expansion. Out of the four models tested, by far the best results are obtained for the Blast-Wave model with resonances and $a=-0.5$. This shows that the hypersurfaces with $\rho$ decreasing with time ({\em cf.} Fig.~\ref{fig:models}) are favored. Note that the $R_{\rm long}$ radius is particularly sensitive to the $a$ parameter, with $a=-0.5$ giving the right result, while increasing $a$ spoils the agreement. The model values of the intercept $\lambda$ shown in Figs.~\ref{fig:cra}-\ref{fig:bwm} are too large compared to the data, which simply reflects the fact that we do not take into account the effect of secondary pions coming from the weak decays, as well as the contamination of the pion sample by misidentified particles in the experiment. 

The Cracow model and the Blast-Wave model with $a=0.5$ have very close predictions, as expected from the similarity of the hypersurfaces, {\em cf.} Fig.~\ref{fig:models}. We note that in all considered models the qualitative behavior of the dependence of the radii on $k_T$ is correct.  The Coulomb corrections evaluated with the Bowler-Sinyukov formalism are small, of the order of a small fraction of a fermi.

Our method of determining the HBT radii from the model involves the calculation of the complete 3-dimensional correlation function. First we use the free wave function of Eq. (\ref{psiq}) in Eq. (\ref{cfbysum}) and compare it to the gaussian fit made according to Eq. (\ref{cfgaus}). The results are shown in Fig. \ref{fig:cf3dqs} where we plot the projections of the correlation function itself, as well as the projections of the 3-dimensional fit. The deviations between the function (symbols) and the fit (curves) reflect the fact that the underlying two-particle distributions are not gaussian and produce a non-gaussian correlation function. One can also see that the increase of the integration regions in the complementary directions leads to better agreement. Since this is an important finding, we restate this observation: In a fixed $k_T$ bin we take the 3-dimensional function. We choose one of the directions, say $q_{\rm out}$, and integrate over the remaining two directions, $q_{\rm side}$ and $q_{\rm long}$, within the specified range. When the range is very narrow, this corresponds to slicing the 3-dimensional function along the line $q_{\rm side} = q_{\rm long} = 0$. This gives the circles in Fig. \ref{fig:cf3dqs}. Next we repeat the same projection prescription but for the gaussian fit to the correlation function, which results in the solid lines. The lines deviate from the circles within a few percent showing that the gaussian approximation works at that level.  Increasing the integration range in the complementary directions results in a much better agreement, which is manifested by the overlap of the squares to the dotted lines. One can also see that even though the detailed shape of the correlation function is not exactly reproduced when the integration region is smaller (circles and triangles), the overall width, and hence the radius, is described well.  

Now we come to the analysis of the correlation function taking into account the Coulomb interactions. Explicitly,  we use the Coulomb wave function of Eq. (\ref{psiqc}) in Eq. (\ref{cfbysum}) and compare it to the Bowler-Sinyukov formula (\ref{cffitbs}). We perform the same procedure as above and the results are shown in Fig. \ref{fig:cf3dqsc}. We observe that the Coulomb interactions dig holes at low values of $q$, which is the well-known result of the long-range repulsion.

In Fig. \ref{fig:rors} we show the ratio $R_{\rm out}/R_{\rm side}$ for several models. The very good agreement with the STAR data (open circles) is obtained for the Blast-Wave model with resonances and $a=-0.5$. This behavior is already expected from the inspection of Figs.~\ref{fig:cra}-\ref{fig:bwm}. The Cracow model with $T=165$ MeV describes well the $k_T$ dependence of the experimental ratio, giving the magnitude of the ratio smaller by about 15\%.

\section{Effects of resonances in the correlation function}

Figure \ref{fig:res} shows the separation distributions for the Blast-Wave model with $a=-0.5$ and for the bin  0.25 GeV $< k_T <$ 0.35 GeV. On the left-hand-side the distributions of pairs constructed only from the primordial pions are shown. It can be seen that these distributions are cut off at the value determined by the $\rho_{\rm max}$ parameter of the Blast-Wave model, since by definition no pair can have a separation greater than $2\rho_{\rm max}$ in the {\it out} and {\it side} directions. The correlation function is the Fourier transform of these distributions. 
As described in Sect. \ref{getrad}, the correlation function may be fitted with the gaussian formula (\ref{cfgaus}) and the result of the fit can be used to find the presumed gaussian distribution (\ref{staticsource}).  The separation distributions obtained with the help of such a procedure, i.e., from the gaussian fits to the correlation function, are shown as thin lines in Fig. \ref{fig:res}. We note that the gaussian approximation works reasonably well at low values of the separation radii with some deviations in the case of long direction at large values of $r_{\rm long}$. In that case a tail results from the chosen model of expansion. In our boost-invariant model the distribution extends to infinity, however, the observed drop is caused by the presence of the homogeneity length in the system.

\begin{figure*}[t!]
\begin{center}
\includegraphics[width=0.8\textwidth]{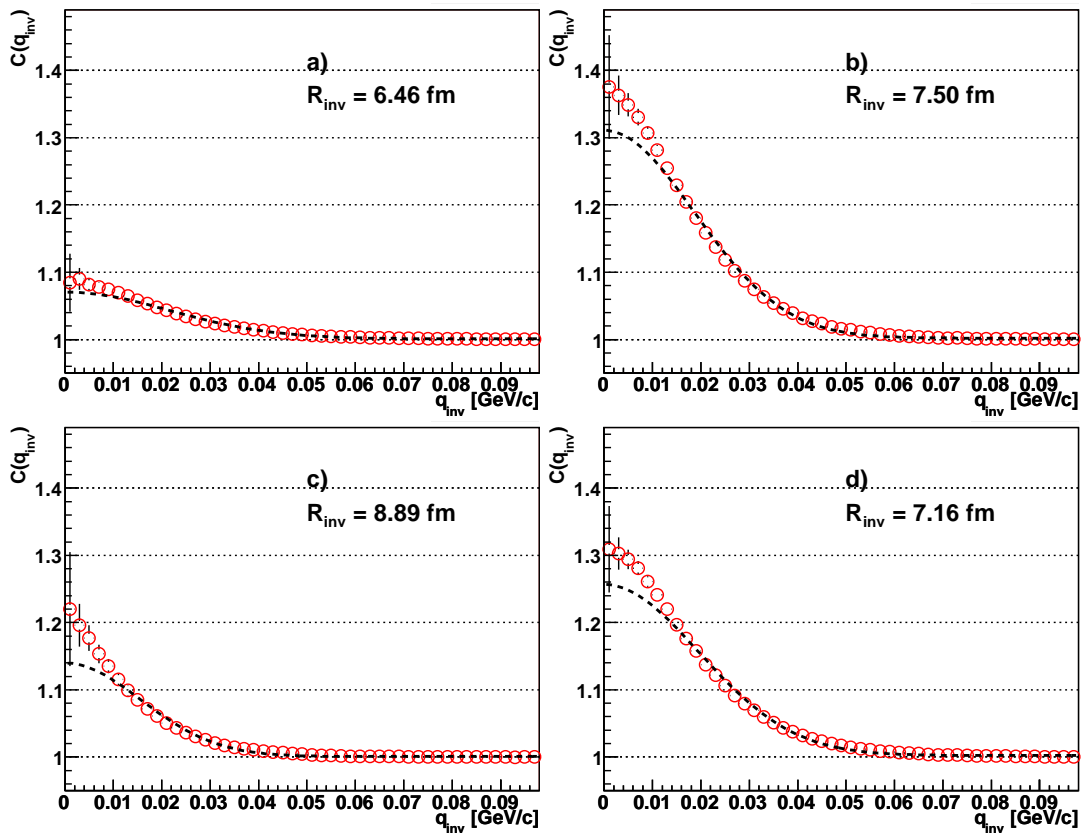}
\end{center}
\caption{The 1-dimensional correlation function plotted as a function of $q_{\rm inv}$ for the Blast-Wave model with resonances, $a=-0.5$, 0.25 GeV $ < k_T < $ 0.35 GeV (open circles). Panel a) gives the "primary-primary" case, b) - "primary, other - $\rho$", c) - "primary, $\rho$, other - $\omega$", and d) - "other, primary - other". For comparison dashed lines show the best fit of the gaussian formula, with the corresponding values of $R_{\rm inv}$.}
\label{fig:cfanatom}
\end{figure*}

On the right-hand-side of Fig. \ref{fig:res} we show the corresponding plots including all pairs of the pions created before $500$ fm/c, i.e., the primordial pions and other pions coming from almost all strongly-decaying resonances. The effect of the resonances is very clearly visible in a long-range tail in the {\it out} and {\it side} directions. One can see how the model curve departs from the gaussian fit around 17 fm. For the {\it long} direction the tail is also produced and the effect adds up to the tail produced by the expansion model. Thus, it is clear that the
gaussian fit to the correlation functions has no way to describe the long-range tail of the distributions. The tails have, however, an effect on the correlation function and show up as peak at low $q_{\rm inv}$ which is seen in Fig.\ref{fig:cf3dqs} as the difference between the data (circles) and fit (solid lines). This results in the lowering of the $\lambda$ parameter of the gaussian fit \cite{Baym:1997ce}.

It is interesting to study the long-range tails of the separation distributions in more detail. Figure \ref{fig:anatom} shows the anatomy of the separation distributions divided into several components. The pions are divided into four groups: primary, those coming from $\rho$ decays, those coming from $\omega$ decays, and other coming from decays of other resonances than $\rho$ or $\omega$. In the plot we show the distributions of pairs constructed from pions belonging to the classes defined above. First we consider pairs where both pions are primary (circles). 
One can see that in the case of {\it out} and {\it side} directions these pions are concentrated near the origin with $2 \rho_{\rm max} \approx$~18~fm providing the cutoff. There is no such cutoff in the long direction where we can see the falloff resulting from the homogeneity length. Next we consider the pairs where one of the pions comes from the $\rho$ decays and the other from primary pions or pions coming from decays of  resonances other than $\omega$ and $\rho$ (up-triangles, labeled as "primary, other - $\rho$"). These pairs are responsible for the increase of the strength of the source in all three directions. In the {\it out} and {\it side} directions, they cause the swelling of the source. The curves corresponding to the pairs where one of the pion belongs to the group "other" and the second to the groups "other, primary" (down-triangles) show a very similar behavior to the "primary, other, $\rho$" case. Finally, we show the pairs where one of the pions comes from the $\omega$ decay (squares). In all three directions we observe the long-range tails caused by the long lifetime.  These pairs produce a non-gaussian character of the correlation function. It can also be seen, that the total distribution (solid lines) can be well approximated by a combination of a gaussian-like core at low $r$ and the long-range non-gaussian halo. 

In order to illustrate the influence of the resonances on the correlation function itself, in Fig. \ref{fig:cfanatom} we show how different types of pairs contribute to the observed correlation as a function of $q_{\rm inv}$. It is clearly seen that none of the contributions is well-reproduced by a gaussian. This feature is most prominently seen for the pairs which contain at least one pion from the $\omega$ decay: in this case the correlation function has a peak at low $q_{inv}$, which is expected as the spacetime distribution for these pairs, seen in Fig. \ref{fig:anatom} is exponential at large $r$. Even though the gaussian fit fails to describe the detailed behaviour of the functions, especially at low values of $q_{\rm inv}$, it serves as a tool for extraction of the HBT radii. One can see, that the smallest radius is obtained for the primordial pairs, as expected. However, they provide only about 10\% of the correlation effect. Pairs, which contain a pion from any strongly-decaying resonance, except for $\rho$ or $\omega$, show a size larger by approximately 1 fm. They account for at least 30\% of the correlation effect. Pairs containing the $\rho$-decay product show slightly larger increase of the radius (also about 1 fm) and provide the largest contribution to the correlation effect, about 40\%. Pairs containing the $\omega$-decay products give the largest size, as expected, but their contribution to the correlation function is sharply peaked at small $q_{\rm inv}$, which results in the decrease of the $\lambda$ parameter, but does not influence the width of the correlation function (and therefore the obtained radius).

\begin{figure}[t]
\begin{center}
\includegraphics[angle=0,width=0.43\textwidth]{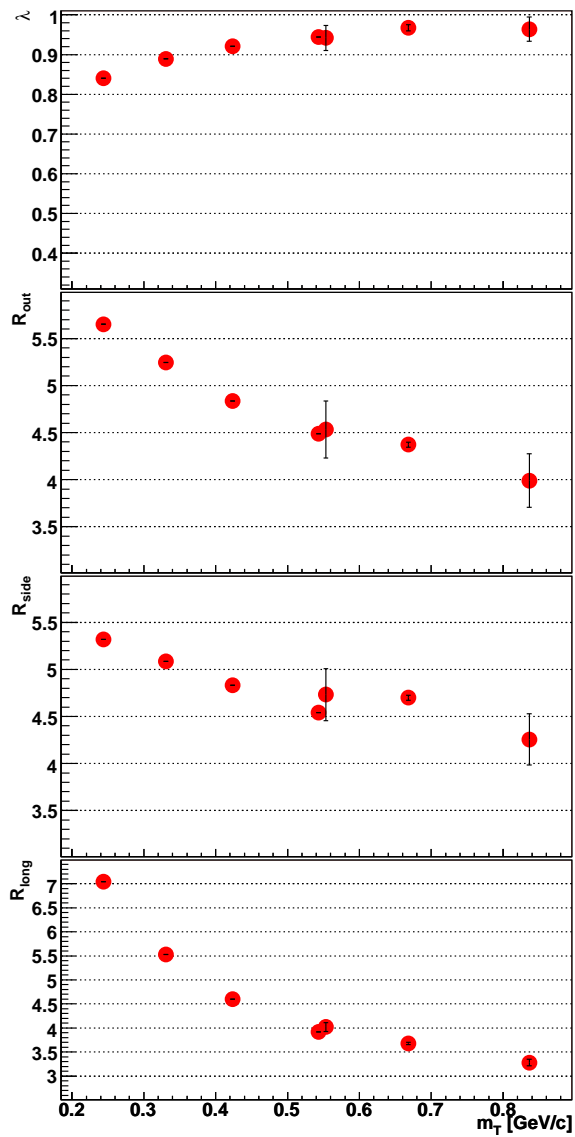}
\end{center}
\caption{Predictions for the kaon femtoscopy in the Blast-Wave model with $a=-0.5$.
The model parameters are $T=165.6$~MeV, $\mu_B=28.5$~MeV, $\tau=8.55$~fm,
$\rho_{\rm max}=8.92$~fm, $v_{T}=0.311$~c. The first four points
from the left are predictions for pions, the three points on the right
- for kaons.}
\label{fig:kaons}
\end{figure}

\section{Predictions for the kaon}

In this section we show our predictions for the kaon HBT radii from the Blast-Wave model with resonances. The results obtained with the free wave function (\ref{psiq}) and formula (\ref{cfbysum}) are shown in Fig.~\ref{fig:kaons}. We also give the results for the pions and notice that the kaons nicely extend the curves to higher values of the transverse mass. The results are plotted as a function of $m_{T}$ of the pair. This choice follows from the expectation that the radii are influenced mostly by the flow, which results in the approximate $m_{T}$ scaling \cite{Pratt:1984su,Akkelin:1996sg}. The error bars indicate the errors of the gaussian fit to the {\tt THERMINATOR} simulation. We note that for all radii the scaling holds within these uncertainties.
The calculation presented in Fig.~\ref{fig:kaons} uses the model which was most successful for the pions. Note that all parameters were fixed by the pionic sector and the predictions for kaons involve no parametric freedom. 

\section{Conclusions}

In this paper, a class of hydro-inspired models with single freeze-out was used 
to analyze in detail the pion correlation functions, with an emphasis on the role of the chosen model of freeze-out as well as the detailed influence of resonance decays. Concerning the choice of the expansion model we have found that the freeze-out geometry where the transverse size decreases with time works by far the best, allowing  for a uniform description of the $p_T$ spectra and the HBT correlation radii. Such a model has the features similar to those found in the advanced hydro calculations, however in our study the resulting time and size scales are shorter. 
We have found that the $R_{\rm long}$ radius is particularly sensitive to the details of expansion, which helps to discriminate between various cases. Our calculations were done for the Cracow single-freeze-out model as well as for the Blast-Wave model including all resonance decays, and with a modified shape of the $\rho-t$ freeze-out curve. We have achieved a satisfactory and uniform agreement for the description of the data, as
can be judged from Figs. 3 and 7. 

Our calculations used the code {\tt THERMINATOR}, which is a Monte-Carlo implementation of the hydro-inspired models with single freeze-out. The use of the Monte-Carlo technique allowed us to study in a greater detail the effects of resonances on the shape of the correlation functions. Since our freeze-out temperature is large, we need to include practically all resonances, as in the studies of particle abundances and momentum spectra. We have explicitly found non-gaussian features of the pion correlation functions caused by the long-living resonances, mainly the $\omega$ meson, confirming earlier expectations. In addition, we have carefully discussed their quantitative role for the extraction and interpretation of the HBT radii, as well as the shape of the correlation functions and the separation distributions. In short, we hope that our analysis provides a very useful ``vivisection'' of the pion HBT problem, helpful in the understanding of the underlying space-time picture of relativistic heavy-ion collisions. We find that the pion HBT data from RHIC are fully compatible with the single freeze-out scenario. 

Finally, we give predictions for the HBT radii of kaons, which should be measured shortly. The results for the kaons exhibit the $m_T$-scaling proposed in Ref. \cite{Pratt:1984su,Akkelin:1996sg}.

\appendix

\section{Correlation functions for primordial particles}

In the case of the primordial particles one may obtain analytic expressions for the correlation functions which are very useful as the reference point for the Monte-Carlo method.  They can also be used for testing numerical calculations.

\subsection{Side correlation function}
In the case of the side correlation function, one may choose the direction of the average transverse momentum 
to be parallel to the $r_x$ - axis and the difference of the two momenta to be parallel to the $r_y$ - axis. 
In this reference frame, at zero rapidity ($y=0$) we have
\begin{equation}
{\vec k} = (k_\perp,0,0), \quad {\vec q} = (0,q_{\rm side}\, ,0),
\nonumber
\label{kvectside}
\end{equation}
and the phase of the Fourier transform of the emission function has the form
\begin{equation}
i\, q \cdot x = -i\, q_{\rm side}\, y = -i\,  q_{\rm side}\, \rho \sin\phi.
\nonumber
\label{bmqxside}
\end{equation}
Performing similar steps as those which led us to Eq. (\ref{dNdydptKI}) we obtain the Fourier transform of 
the emission function in the following form $(q=q_{\rm side})$
\begin{eqnarray}
&& S_{\rm side}(k_\perp,q) =  {1 \over 2\pi^2} \sum\limits_{n=1}^\infty (\mp)^{n+1}
e^{n \beta \mu}
\int\limits_0^1 d\zeta \,\, \rho(\zeta) {\tilde \tau}(\zeta) \nonumber \\
&& \times \left\{
m_\perp {d\rho \over d\zeta} 
K_1 \left[n \beta m_\perp \hbox{cosh}\alpha_\perp \right]
{\cal I}_0 \left[n \beta k_\perp \hbox{sinh}\alpha_\perp,q \rho(\zeta) \right] \right. \nonumber \\
&& \left. - p_\perp { d{\tilde \tau} \over d\zeta} 
K_0 \left[n \beta m_\perp \hbox{cosh}\alpha_\perp \right]
{\cal I}_1 \left[n \beta k_\perp \hbox{sinh}\alpha_\perp,q \rho(\zeta) \right]
\right\}. \nonumber \\
\nonumber
\label{primside}
\end{eqnarray}
Here we introduced two functions:
\begin{equation}
{\cal I}_0(a,b) = {1 \over 2\pi} \int\limits_0^{2\pi}
d\phi \exp(a \cos\phi-i b \sin\phi),
\nonumber
\label{calI0}
\end{equation}
and
\begin{equation}
{\cal I}_1(a,b) = 
{1 \over 2\pi} \int\limits_0^{2\pi} d\phi \, \cos\phi \,
\exp(a \cos\phi-i b \sin\phi).
\nonumber
\label{calI1}
\end{equation}
For $b=0$ they are reduced to the Bessel functions $I_0(a)$ and $I_1(a)$. 

\subsection{Out correlation function}
In this case we may choose the coordinate system where
\begin{eqnarray}
{\vec k} = (k_\perp,0,0), \quad {\vec q} = (q_{\rm out}\, ,0,0),
\nonumber
\label{kvectout}
\end{eqnarray}
and the phase of the Fourier transform is
\begin{eqnarray}
i\, q \cdot x &=& i\, q^0  t - i\, q_{\rm side}\,  x \nonumber \\
&=& i\, q^0 \, {\tilde \tau} \, \hbox{cosh}\alpha_\parallel 
- i\, q_{\rm side}\,  \,\rho \,\cos \phi.
\nonumber
\label{bmqxout}
\end{eqnarray}
The energy difference $q^0$ in Eq. (\ref{bmqxout}) is obtained from the
formula
\begin{eqnarray}
q^0 = \sqrt{m^2 + (k_\perp + q_{\rm side}\, /2)^2} - \sqrt{m^2 + (k_\perp - q_{\rm side}\, /2)^2}, \nonumber \\
\nonumber
\end{eqnarray}
and the Fourier transform of the emission function reads $(q=q_{\rm side})$ 
\begin{widetext}
\begin{eqnarray}
S_{\rm out}(k_\perp,q) &=&  
{1 \over 2\pi^2} \sum\limits_{n=1}^\infty (\mp)^{n+1} e^{n \beta \mu}
\int\limits_0^1 d\zeta \,\, \rho(\zeta) {\tilde \tau}(\zeta)  \left\{
m_\perp {d\rho \over d\zeta} 
K_1 \left[n \beta m_\perp \hbox{cosh}\alpha_\perp\!-i q^0 \, {\tilde \tau} \right]
I_0 \left[n \beta k_\perp \hbox{sinh}\alpha_\perp\!-i q \rho \right] \right.
\nonumber \\
&& \left. - p_\perp { d{\tilde \tau} \over d\zeta} 
K_0 \left[n \beta m_\perp \hbox{cosh}\alpha_\perp\!-i q^0 \, {\tilde \tau} \right]
I_1 \left[n \beta k_\perp \hbox{sinh}\alpha_\perp\!-i q \rho \right]
\right\}. \nonumber \\
\nonumber
\label{primout}
\end{eqnarray}
\end{widetext}

\subsection{Long correlation function}
Similarly to the two previous cases, we find
\begin{equation}
{\vec k} = (k_\perp,0,0), \quad {\vec q} = (0,0,q_{\rm long}\, ),
\nonumber
\label{kveclong}
\end{equation}
\begin{equation}
i\, q \cdot x = -i\, q_{\rm long}\,  z  = -i\, q_{\rm long}\,  
{\tilde \tau} \hbox{sinh}\alpha_\parallel,
\nonumber
\label{bmqxlong}
\end{equation}
and $(q=q_{\rm long})$ 
\begin{eqnarray}
&& S_{\rm long}(k_\perp,q) = {1 \over 2\pi^2} \sum\limits_{n=1}^\infty (\mp)^{n+1}
e^{n \beta \mu} 
\int\limits_0^1 d\zeta \,\, \rho(\zeta) {\tilde \tau} (\zeta) \nonumber \\
&& \times \left\{
m_\perp {d\rho \over d\zeta} 
{\cal K}_1 \left[n \beta m_\perp \hbox{cosh}\alpha_\perp, q {\tilde \tau} \right]
I_0 \left[n \beta k_\perp \hbox{sinh}\alpha_\perp\right] \right.
\nonumber \\
&& \left. - p_\perp { d{\tilde \tau} \over d\zeta} 
{\cal K}_0 \left[n \beta m_\perp \hbox{cosh}\alpha_\perp, q {\tilde \tau} \right]
I_1 \left[n \beta k_\perp \hbox{sinh}\alpha_\perp\right]
\right\}. \nonumber \\
\nonumber
\label{primlong}
\end{eqnarray}
The functions ${\cal K}_0$ and ${\cal K}_1$ are defined by the integrals:
\begin{equation}
{\cal K}_0(a,b) = {1\over 2} \int\limits_{-\infty}^{\infty}
d\alpha  \exp(- a \hbox{cosh}\alpha-i b \hbox{sinh}\alpha)
\nonumber
\label{calK0}
\end{equation}
and
\begin{equation}
{\cal K}_1(a,b) = {1\over 2} \int\limits_{-\infty}^{\infty}
d\alpha \,  \hbox{cosh}\alpha
\exp(- a \hbox{cosh}\alpha-i b \hbox{sinh}\alpha).
\nonumber
\label{calK1}
\end{equation}
For $b=0$ these functions reduce to the modified Bessel functions $K_0$ and $K_1$.

\bibliography{femto11}

\end{document}